\documentclass[12pt]{article}         
\usepackage{epsfig}                   
\newfont{\ninerm}{cmr9}
\newcommand{\h}{\hspace*{5 ex}}
\def\dfrac{\displaystyle\frac}
\def\pmb#1{\setbox0=\hbox{#1}%
  \kern-.025em\copy0\kern-\wd0
  \kern.05em\copy0\kern-\wd0
  \kern-.025em\raise.0433em\box0 }

\def\alfa{4\pi\alpha}
\def\g2r2{\frac{G}{2\sqrt{2}}}
\def\geff{g_{eff}}
\def\gve{g_V^e}
\def\gae{g_A^e}
\def\sfi{sin \phi}
\def\cfi{cos \phi}
\def\ss2fi{sin 2 \phi}
\def\cc2fi{cos 2 \phi}
\def\GAS{G_A^{(s)}(Q^2)}
\def\GES{G_E^{(s)}(Q^2)}
\def\GMS{G_M^{(s)}(Q^2)}
\def\GAT0{{\tilde G}_A^{T=0}(Q^2)}
\def\GATUNO{{\tilde G}_A^{T=1}(Q^2)}
\def\etas{\eta_s}
\def\mus{\mu_s}
\def\rhos{\rho_s}
\def\rdues{r^2_s}
\def\RAT1{R_A^{T=1}}
\def\RATZERO{R_A^{T=0}}
\def\RAS{R_A^{(0)}}
\def\laes{\lambda_E^{(s)}}
\def\lams{\lambda_M^{(s)}}
\def\laas{\lambda_A^{(s)}}
\def\j5{j_5}
\def\js5{j_5^+}
\def\Jem{J^{(em)}}
\def\Jnc{J^{(NC)}}
\def\Jems{J^{(em)+}}
\def\Jncs{J^{(NC)+}}
\def\Jncv{J^{(NC)V}}
\def\Jnca{J^{(NC)A}}
\def\ela{\epsilon_{(\lambda)}}
\def\elas{\epsilon_{(\lambda)}^*}
\def\elamus{\epsilon_{(\lambda)}^{\mu *}}
\def\elap{\epsilon_{(\lambda^\prime)}}
\def\elaps{\epsilon_{(\lambda^\prime)}^*}

\def\dsigma{\frac
    {d^3\sigma}{dE_{e^{\prime}}d\Omega_{e^{\prime}}d\Omega_{N}^{c.m.}}} 

\def\tetacm{\vartheta_{c.m.}}
\def\tetap{\tetacm^p}
\def\tetan{\tetacm^n}
\def\Apteta{{\cal A}_p(\tetacm)}

\def\Ateta{{\cal A}(\tetacm)}
\def\Ap{{\cal A}_p}
\def\An{{\cal A}_n}
\def\Ad{{\cal A}_d}
\def\APC{{\cal A}_{PC}(\tetap)}
\def\q2{q^2}
\def\qq2{ \frac {Q^2}{\qlab^2} }

\def\qcm{q_{c.m.}}
\def\qlab{q_{lab}}

\def\tem{t^{(em)}_{s m_s \lambda m_d}}

\def\temsllp{t^{(em)*}_{s m_s \lambda^{\prime} m_d}}

\def\tncv{t^{(NC)V}_{s m_s \lambda m_d}}
\def\tncvsllp{t^{(NC)V*}_{s m_s \lambda^{\prime} m_d}}
\def\tnca{t^{(NC)A}_{s m_s \lambda m_d}}
\def\tncasllp{t^{(NC)A*}_{s m_s \lambda^{\prime} m_d}}
\def\tmatrix{t_{s m_s \lambda m_d}}
\def\tg2{\tan^2{ \frac {\vartheta_{e'}}{2} } }
\def\t2g{\tan^2(\vartheta_{e'}/2)}
\def\c2t{\cos^2(\vartheta_{e'}/2)}
\def\vllp{v_{\lambda\lambda^{\prime}}}
\def\vvllp{v^{(VV)}_{\lambda\lambda^{\prime}}}
\def\vallp{v^{(VA)}_{\lambda\lambda^{\prime}}}

\def\vv0llp{v^{(VV)0}_{\lambda\lambda^{\prime}}}
\def\vvhllp{v^{(VV)h}_{\lambda\lambda^{\prime}}}
\def\va0llp{v^{(VA)0}_{\lambda\lambda^{\prime}}}
\def\vahllp{v^{(VA)h}_{\lambda\lambda^{\prime}}}
\def\vl0{v^0_L}
\def\vt0{v^0_T}
\def\vtl0{v^0_{TL}}
\def\vtt0{v^0_{TT}}
\def\vht{v^h_T}
\def\vhtl{v^h_{TL}}
\def\dellp{ \left( \frac {1+\delta_{\lambda+\lambda^{\prime},1}}  
                  {1+\delta_{\lambda,0}} \right)}
\def\fem#1{f^{(em)}_{#1}}
\def\fhemtl{f^{(em)h}_{TL}}
\def\femv#1{f^{(em-V)}_{#1}}
\def\fhemvtl{f^{(em-V)h}_{TL}}
\def\fema#1{f^{(em-A)}_{#1}}
\def\fhematl{f^{(em-A)h}_{TL}}
\def\fhemat{f^{(em-A)h}_{T}}
\def\femllp{f^{(em)}_{\lambda\lambda^\prime}}
\def\fhemllp{f^{(em)h}_{\lambda\lambda^\prime}}
\def\femvllp{f^{(em-V)}_{\lambda\lambda^\prime}}
\def\femallp{f^{(em-A)}_{\lambda\lambda^\prime}}
\def\fhemvllp{f^{(em-V)h}_{\lambda\lambda^\prime}}
\def\fhemallp{f^{(em-A)h}_{\lambda\lambda^\prime}}
\def\wemllp{w^{(em)}_{\lambda\lambda^\prime}}
\def\wemvllp{w^{(em-V)}_{\lambda\lambda^\prime}}
\def\wemallp{w^{(em-A)}_{\lambda\lambda^\prime}}
\def\INFN{ Istituto Nazionale di Fisica Nucleare, Sezione di Firenze}
\def\Via{I-50125 Firenze, Italy}
\def\Dipa{ Dipartimento di Fisica, Universit\`a di Firenze}
\title{
{\bf Deuteron Electroweak Disintegration} }

\author{B. Mosconi\\                    
                {\small  \Dipa } \\
                {\small  \INFN } \\
                {\small  \Via}   \\
and \\
P. Ricci\\
                {\small  \INFN } \\
                {\small  \Via }
}
\date{February 25, 1997}
\begin{document}

\baselineskip=20pt
\maketitle
\thispagestyle{empty}
\begin{abstract}
We study the deuteron electrodisintegration with inclusion of the neutral 
currents focusing on the helicity asymmetry of the exclusive cross 
section in coplanar geometry  $\Ateta$.
We stress that a measurement of $\Ateta$ in the quasi elastic region 
is of interest for an experimental determination of the weak form 
factors of the nucleon, allowing one to obtain the parity violating 
electron neutron asymmetry. Numerically, we consider the reaction at 
low momentum transfer and discuss the sensitivity of $\Ateta$ to the 
strangeness radius and magnetic moment. The problems coming from the 
finite angular acceptance of the spectrometers are also considered. 
\end{abstract}

\

\

PACS number(s):  24.80.+y , 14.20.Dh , 25.10.+s , 25.30.Fj 
\newpage
\section{Introduction}
\h Parity violating (PV) electron scattering probes weak neutral currents 
and can provide very interesting information on the strange-quark 
contributions to the electroweak (ewk) form factors of the nucleon 
and on the weak coupling constants at the hadronic level. 
Since different theoretical models give largely different predictions 
for the strange vector ($G_E^s(Q^2)$, $G_M^s(Q^2)$) 
and axial-vector ($G_A^s(Q^2)$) 
form factors as well as for the radiative corrections to the weak coupling 
constants, one has to make recourse to an experimental determination 
of these quantities. 
For this, one needs to isolate observables which are selectively 
sensitive to one or the other unknown quantity. 
It will take a number of measurements in neutrino scattering, PV atomic  
experiment and PV electron scattering 
to determine these form factors and coupling constants. 
The best information on $G_A^s(Q^2)$ is expected from elastic 
neutrino scattering experiments where theoretical uncertainties in 
higher - order processes are small. 
The BNL experiment 734 \cite{Ahrens87} 
already determined a non-zero $G_A^{(s)}(0)$ 
even if with large errors \cite{Garvey93}. 
Results of the spin-dependent deep inelastic lepton scattering 
experiments off protons \cite{Ashman8889,Adams94,Abe95p} and off neutrons 
\cite{Adeva93,Adams95,Anthony93,Abe95n} 
confirm such finding, again with large theoretical errors because 
of the application of SU(3) flavor symmetry to hyperon decays. 
The LSND experiment on neutrino oscillations \cite{LSND89} presently 
underway at LAMPF should better constrain $G_A^{(s)}(0)$. 
The suggestion that the strangeness magnetic moment $\mus = G_M^s(0)$ 
could be determined measuring  
the  PV asymmetry in elastic $\vec{e} p$ scattering at backward angles 
was put forward by McKeown \cite{McKeown89} and Beck \cite{Beck89}. 
A first experiment \cite{SAMPLE96} aiming to place limits on $\mus$ 
is already underway at Bates Laboratory. 
Measurements at forward angles could be used to constrain $G_E^s(Q^2)$. 
The accuracy of such experiments using only a proton target is strongly 
limited, because of the complications from radiative corrections 
\cite{MusolfPLB90} to the dominant isovector axial-vector coupling. 
Measuring PV asymmetry in electron scattering from nuclei, 
where different isospin combinations can be realised, 
seems a promising way-out to disentangle radiative corrections and 
strange-quark contributions. 
In particular, the 
PV electron scattering from isoscalar and spinless nuclei, such as 
$^4He$, where only the electric weak current can contribute, could 
lead to a determination of $G_E^s(Q^2)$ \cite{MusolfPRC94}. 
Two experiments of PV electron scattering off complex nuclei 
have already been carried out \cite{Souder90,Heil89} and 
several others are in preparation at Bates, CEBAF 
and MAMI. For a review we refer to the paper 
by Musolf et al.\cite{MusolfPR94} who present 
a very detailed discussion 
of the intermediate-energy semileptonic probes of the hadronic neutral 
current. 
Different theoretical approaches have been pursued going from the 
relativistic Fermi gas model \cite{Donnelly92} to the relativistic 
mean field theory \cite{Horowitz93} to the continuum shell model 
\cite{Amaro96}. 
Also the case of the deuteron has been studied extensively~ 
\cite{Schramm94,Hadjimichael92}. 
\par
Up to now, only the helicity asymmetry of the elastic cross section and 
of the inclusive inelastic cross section in PV  
electron scattering has been 
considered~\cite{MusolfPR94,Donnelly92,Horowitz93}.   
In this paper we study the helicity asymmetry of the cross section for 
the exclusive PV electron deuteron scattering in the in-plane kinematics. 
In general, namely in the out-of-plane geometry, the helicity asymmetry 
is not zero even in the parity conserving (PC) electrodisintegration 
where it is given by the so-called fifth structure function.
Instead, the helicity asymmetry of the in-plane kinematics reaction must 
vanish in a PC theory. This can be seen using simple geometrical 
considerations. In fact, the image of the reaction given by a mirror 
parallel to the scattering plane is the same as the original reaction 
apart from the change of helicity of the incoming electron. Therefore, 
if parity is conserved the two processes proceed with equal probability 
leading to a vanishing asymmetry.
\par
We expect that the 
obvious drawback of the reduced counting rates of the coincidence 
experiments might be compensated by the enhanced sensitivity to 
the form factors of the  nucleon detected in coincidence with the electron. 
In fact, this is the case in the PC electron-deuteron 
scattering at the quasi-elastic (QE) peak. It turns out that the deuteron 
can be confidently used as a quasi-free neutron target in that region. 
Therefore, from measurements of $({\vec e},e^\prime p)$ and 
$({\vec e},e^\prime n)$ 
reactions it should be possible to get information on the isoscalar 
form factors which take contributions from the strange quark. 
\par
We shall neglect the effects of the PV nuclear interactions. 
In fact, previous studies have shown these PV 
effects to be small in deuteron photodisintegration~\cite{Hadjimichael71} 
as well as in elastic and inelastic electron deuteron 
scattering~\cite{Hwang8081} 
except for very low-energy electrons. 
\par 
In Sect.2 we describe our treatment of the 
PV e-d inelastic scattering and we give the general 
expression of 
the helicity asymmetry of the 
coincidence cross section $\Apteta$. 
We also discuss its sensitivity to the weak nucleon form factors. 
In Sect.3 we present our 
numerical results for the exclusive asymmetry in 
QE kinematics at Q$^2$ = 0.1 (GeV)$^{-2}$.
Finally, in Sect.4 we state our conclusions. 
\section{Formalism}
\subsection{Parity-violating exclusive cross section}
\h The invariant amplitude for 
the parity-violating exclusive deuteron electrodisintegration, 
to lowest order, is the sum of the 
one-photon and the one-$Z^0$ boson exchange process: 
\begin{equation}
  {\cal{M}} = {\cal{M}}_{[\gamma]} + {\cal{M}}_{[Z^0]} ~~~~~~,\label{e:M} 
\end{equation}
with
\begin{equation}
  {\cal{M}}_{[\gamma]} = -\alfa ~j_\mu D^{\mu\nu}_{[\gamma]}(Q^2) ~\Jem_\nu 
                                         ~~~~~~, \label{e:Mgamma}
\end{equation}
and
\begin{equation}
  {\cal{M}}_{[Z^0]} = \g2r2 M^2_Z (\gve j_\mu + \gae  j_{\mu 5}) 
            ~D^{\mu\nu}_{[Z^0]}(Q^2) ~\Jnc_\nu ~~~~~~, \label{e:Mzeta}
\end{equation}
where $Q^2 = - q_\mu^2 > 0$ is the four momentum 
transfer squared; 
$\alpha$ is the fine-structure constant; $G$ is the weak Fermi constant; 
$M_Z$ is the $Z^0$ mass; $\gve$ and $\gae$ are 
the neutral vector and axial-vector couplings of the electron which, in the 
Standard Model, are given by $\gae = 1$ , 
$\gve = -1+4 sin^2\vartheta_W \simeq -0.092$, 
$\vartheta_W$ being the Weinberg or weak-mixing angle. 
The conventions of Musolf et al. \cite{MusolfPR94} for the weak coupling
constants are assumed. 
\par
The electron vector and axial-vector currents are given by the Dirac form
\begin{eqnarray}
          j_\mu &=& \bar u(k^\prime,s^\prime) \gamma_\mu u(k,s) 
                                   ~~~~~,\nonumber \\
      j_{\mu 5} &=& \bar u(k^\prime,s^\prime) \gamma_\mu \gamma_5 u(k,s) 
                                   ~~~~~~, \label{e:jmu}
\end{eqnarray}
where $u(k,s)$ is the electron spinor,  $(k,s)$ and $(k^\prime,s^\prime)$ 
being the four-momentum and spin of the incoming and outgoing electron, 
respectively. 
\par
As for the hadronic currents, $\Jem$ is the electromagnetic (em) current 
and $\Jnc$ the neutral current which consists of a vector and an axial-vector 
component
\begin{equation}
       \Jnc = \Jncv + \Jnca ~~~~~~~.\label{e:JNC}
\end{equation}
\par
Finally, the photon propagator is given by
\begin{equation}
  D^{\mu\nu}_{[\gamma]}(Q^2) = {\frac{1}{Q^2}} (g^{\mu\nu} + 
                               {\frac{q^\mu q^\nu}{Q^2}}) 
                               ~~~~~~,\label{e:Dgam}
\end{equation}
in the Landau gauge, while the $Z_0$ propagator
\begin{equation}
   D^{\mu\nu}_{[Z^0]}(Q^2)= { \frac{g^{\mu\nu} - q^\mu q^\nu / M_{Z}^2}
                            {Q^2 + M_Z^2} } ~~~~~~,\label{e:Dze}
\end{equation}
in the limit $Q^2 \ll M_Z^2$, which we are interested in, becomes 
\begin{equation}
    D^{\mu\nu}_{[Z^0]}(Q^2) \simeq {\frac {g^{\mu\nu}} {M_Z^2} }
                                           ~~~~~~.\label{e:Dzet}
\end{equation}
\par
It is convenient to rewrite the propagators (\ref{e:Dgam}) and 
(\ref{e:Dzet}) in terms of the three 
polarization vectors $\ela^\mu$ ($\lambda = 0, \pm 1$) with 
the properties
\begin{eqnarray}
                q_\mu \ela^\mu &=&0 ~~~~~~~~~~~~~~~~~~~~~~~~~~~,
                                   \label{e:epsa} \\
  g_{\mu\nu} \elamus \elap^\nu &=& (-1)^\lambda \delta_{\lambda,\lambda^\prime} 
                  ~~~~~~~~~~~~~~~~,\label{e:epsb} \\
  \sum_\lambda (-1)^\lambda \elamus \ela^\nu &=& g^{\mu\nu} 
               + {\frac{q^\mu q^\nu}{Q^2}} 
                  ~~~~~~~~~~~~~~.  \label{e:epsc}
\end{eqnarray}
\par
If the momentum transfer ${\bf q}$ is in the $\hat z$ direction 
the polarization vectors take the form
\begin{eqnarray}
    \epsilon^\mu_{(\pm)} &=& \mp {\frac{1}{\sqrt{2}}} (0,1,\pm i,0) 
                            ~~~~~~~~~,\nonumber \\
    \epsilon^\mu_{(0)} &=&  {\frac{1}{Q}} (q_{lab},0,0,\omega_{lab}) 
                            ~~~~~~~, \label{e:epslab}
\end{eqnarray}
in the laboratory (lab) frame
where $q^\mu = \left( \omega_{lab},0,0,q_{lab} \right)$. 
\par
Using the completeness relation (\ref{e:epsc}), the propagators can be written
\begin{eqnarray}
   D^{\mu\nu}_{[\gamma]}(Q^2) & = & {\frac{1}{Q^2}} 
         \sum_\lambda (-1)^\lambda \elamus \ela^\nu 
         ~~~~~~~~~~~~~~~~,\label{e:Dgamma} \\
   D^{\mu\nu}_{[Z^0]}(Q^2) & = & {\frac{1}{M_Z^2}} 
            \left[\sum_\lambda (-1)^\lambda \elamus \ela^\nu 
           - {\frac{q^\mu q^\nu}{Q^2}}  \right] 
         ~~~~~~~~~, \label{e:Dzeta} 
\end{eqnarray}
and the invariant amplitude becomes
\begin{eqnarray}
           {\cal{M}} =&-& {\frac{\alfa}{Q^2}} \sum_\lambda (-1)^\lambda 
                        (j \cdot \elas)(\Jem \cdot \ela) \nonumber \\
                 &+& \g2r2 \sum_\lambda (-1)^\lambda 
                    \left[\gve (j \cdot \elas) + \gae (\j5 \cdot \elas)\right] 
                     (\Jnc \cdot \ela) \label{e:Minv} \\
                 &-& \g2r2 ~{\frac{1}{Q^2}} ~\gae (\j5 \cdot q) (\Jnca \cdot q) 
                    ~~~~~~~~~~, \nonumber
\end{eqnarray}
where we have applied the continuity equations 
\begin{equation}
(j \cdot q) = (\Jncv \cdot q) = 0 ~~~~~~~~~~~.\label{e:jq}
\end{equation}
\par
Clearly expansions (\ref{e:Dgamma}) and (\ref{e:Dzeta}) have allowed 
us to express the 
scattering amplitude as the sum of products of separately Lorentz invariant 
terms (as done by Dmitrasinovic and Gross ~\cite{Dmitrasinovic89}
in the purely em process). 
In actual calculations we shall evaluate the scalar products involving 
the electron current in the lab system and those involving 
the nuclear current in the center of mass (c.m.) system of 
the outgoing nucleons. 
Of course, the transformation of $\ela^\mu$ from the lab 
frame to the c.m. frame must be taken into account.
\par
The next step is to evaluate $\sum_{s^\prime} |{\cal{M}}|^2$ , where 
$s^\prime$ is the spin of the final electron. 
First of all, we neglect the purely weak component 
terms, completely negligible being $\sim G^2$. Moreover, we 
assume, as usual, the extreme relativistic limit (ERL) for the electron 
($m_e \ll E_e$). It is straightforward to see that in this limit
\begin{equation}
  \sum_{s^\prime} (j \cdot \elas) (\js5 \cdot q) = 0 ~~~~~~. \label{e:jj5} 
\end{equation}
Then, the $\gamma - Z^0$ interference contribution involving the 
last term of Eq.(\ref{e:Minv}) vanishes. This means that the term 
$\sim q^\mu q^\nu$ in the $Z^0$  propagator (\ref{e:Dzeta}) does 
not contribute  
and that the $\gamma$ and $Z^0$ propagators can be expressed through 
the completeness relation (\ref{e:epsc}) satisfied by 
the polarization vectors $\ela^\mu$. 
\par
Therefore we obtain
\begin{eqnarray}
  \sum_{s^\prime} |{\cal{M}}|^2 
       &=&{\frac{4 E_e E_{e^\prime} \c2t}{4 m_e^2}} 
         ~\sum_{\lambda\lambda^\prime} (-1)^{\lambda - \lambda^\prime} 
         \left({\frac{q_{c.m.}}{Q}}\right)^{2-|\lambda|-|\lambda^\prime|} 
         \left({\frac{\alfa}{Q^2}}\right)^2 \nonumber \\
       &\times& \Biggl \{\vvllp 
        (\Jem\cdot\ela) (\Jems\cdot\elaps)  \label{e:MM}  \\ 
       &-& \geff ~(\gve \vvllp + \gae \vallp) \nonumber \\
       &\times& \left[ (\Jem\cdot\ela) (\Jncs\cdot\elaps) + 
                      (\Jnc\cdot\ela) (\Jems\cdot\elaps) \right] \Biggr \}
       ~~, \nonumber
\end{eqnarray}
where 
\begin{equation}
 \geff= {\frac{Q^2}{4\pi\alpha}} {\frac{G}{2 \sqrt{2}}}
            ~~~~,\label{e:geff}
\end{equation}
is the effective 
weak coupling constant determining the magnitude of the PV effects in the 
low and medium $Q^2$ and $\vartheta_{e^\prime}$ is the electron 
lab scattering angle.
\par
The electron tensors  $\vvllp$, $\vallp$ which depend on electron 
kinematic variables only, correspond to the products of vector current - 
vector current and vector current - axial-vector current. 
More precisely, they are defined by
\begin{eqnarray}
    \sum_{s^\prime} (j \cdot \elas) (j^+ \cdot \elap)  
   &=&{\frac{4 E_e E_{e^\prime} \c2t}{4 m_e^2}}
    \left({\frac{q_{c.m.}}{Q}}\right)^{2-|\lambda|-|\lambda^\prime|} \vvllp 
                      \nonumber \\
    \sum_{s^\prime} (j \cdot \elas) (\js5 \cdot \elap) 
   &=&{\frac{4 E_e E_{e^\prime} \c2t}{4 m_e^2}}  
    \left({\frac{q_{c.m.}}{Q}}\right)^{2-|\lambda|-|\lambda^\prime|} \vallp 
                ~~~~~~.\label{e:jele}
\end{eqnarray}
\par
It is straightforward to obtain from (\ref{e:MM}) the 
expression of the parity-violating exclusive 
deuteron electrodisintegration cross section for polarized electron beam. 
In terms of the transition matrix elements it reads 
\begin{eqnarray}
\dsigma   &=&{\frac{1}{3}} {\frac{\sigma_M}{M_d}} \sum_{\lambda\lambda^\prime} 
               \sum_{s m_s m_d} 
               \Bigg\{ \vvllp T^{(em)}_{s m_s \lambda m_d} 
               T^{(em)*}_{s m_s \lambda^\prime m_d} \nonumber \\
          &-& \geff ~(\gve \vvllp + \gae \vallp)        \label{e:dsigmaTT} \\ 
          &\times&  \Big[ T^{(em)}_{s m_s \lambda m_d} 
                         T^{(NC)*}_{s m_s \lambda m_d}  
                + T^{(NC)}_{s m_s \lambda m_d} 
                  T^{(em)*}_{s m_s \lambda^\prime m_d} 
                  \Big] \Bigg \} 
           ~~~~~~, \nonumber
\end{eqnarray}
where $\sigma_M$ is the Mott cross section and $M_d$ is the deuteron mass.
The superscripts $(em)$ and $(NC)$  
indicate to which particular nuclear current the $T$-matrix element 
refers to. 
\par
The $T$-matrix elements are related to the hadronic current matrix elements: 
\begin{eqnarray}T_{s m_s \lambda m_d}=
        &-&\sqrt{ \frac{p_{c.m.}E^N_{c.m.}E^d_{c.m.}}
                                   {16 \pi^3}} \nonumber \\
        &\times& (-1)^\lambda \left({\frac{\qcm}{Q}}\right)^{1-|\lambda|}
         \langle s m_s \vert \hat{J} \cdot \ela \vert m_d \rangle
         ~~~~~,\label{e:T}
\end{eqnarray} 
where $\hat{J}$ is the hadronic current operator and 
the nuclear states are defined in the usual non-covariant 
normalization; namely $\vert m_d \rangle$ is the deuteron state
normalized to one, with spin projection $m_d$ on the momentum transfer, 
while the final $np$ state $\vert s m_s \rangle$, 
characterized by spin $s$ and its projection $m_s$ on the relative 
momentum ${\bf p}_{c.m.}$, is normalized so that it becomes 
\begin{equation}
\vert s m_s \rangle = e^{i{\bf p}_{cm}\cdot{\bf r}} \chi_{s m_s} 
                      ~~~~~,\label{e:sms}
\end{equation}
in plane wave (PW) approximation.  
Of course, in order to calculate the matrix elements in Eq.(\ref{e:T}), 
the same quantization axis has to be taken for both initial and final 
states. This simply amounts to the rotation leading ${\bf q}$ into 
${\bf p}_{c.m.}$ or vice versa.
Finally, $E^N_{c.m.}$ and $E^d_{c.m.}$ are the nucleon and 
deuteron c.m. energy, respectively. Note that, 
owing to the factorization of $(\sigma_M/M_d)$ in Eq.(\ref{e:dsigmaTT}), 
the $T$-matrix is dimensionless as that introduced in Ref.\cite{MR90}. 
\par
Further, we remark that the spherical component $\lambda=0$ 
of the nuclear current, given in the c.m. frame by
\begin{equation}
  J \cdot \epsilon_{(0)} = {\frac{\qcm}{Q}} \rho({\bf q}) - 
                 {\frac{\omega_{c.m.}}{Q}} ({\bf J \cdot \hat q}) 
                  ~~~~,\label{e:Jsph}
\end{equation}
can be conveniently written in the case of the em current and of the 
vector component of the neutral current by means of the charge density as 
\begin{equation}
J \cdot \epsilon_{(0)}  = 
              \left({\frac{Q}{\qcm}}\right) \rho({\bf q}) 
               ~~~~,\label{e:Jeps}
\end{equation}
by using the continuity equation to express $({\bf J \cdot \hat q})$ 
in terms of $\rho({\bf q})$. 
\par
In the ERL the electron beam may  
only have longitudinal polarization of degree $h$. 
Therefore, both the electron tensors $\vllp^{(VV,VA)}$ consist of two terms, 
$\vllp^{(VV,VA)}=\vllp^{(VV,VA)0} +h \vllp^{(VV,VA)h}$ which 
correspond to unpolarized  and polarized electrons, respectively. 
\par
It is easy to show that 
$\vallp$ are related to the $\vvllp$ in the following way 
\begin{eqnarray}
    \va0llp &=& \vvhllp ~~~~~~, \nonumber \\
    \vahllp &=& \vv0llp ~~~~~~.    \label{e:vVVA}
\end{eqnarray} 
\par
Of course, the kinematic functions $\vvllp$ coincide with those ($\vllp$) 
appearing in parity conserving  electron scattering  ~\cite{MR90,MPR93}. 
Then, from now on we shall omit any superscript in writing these 
kinematical functions. 
We recall that they  are symmetric 
and satisfy the relations 
\begin{eqnarray}
   \vllp &=&v_{\lambda^{\prime}\lambda} ~~~~~~~~~~,  \nonumber \\
   v^0_{-\lambda-\lambda^{\prime}}
          &=&(-)^{\lambda+\lambda^{\prime}}  
               v^0_{\lambda\lambda^{\prime}}
                                                    ~~~~~, \label{e:vprop} \\ 
   v^h_{-\lambda-\lambda^{\prime}}
         &=&(-)^{\lambda+\lambda^{\prime}+1} 
              v^h_{\lambda\lambda^{\prime}} 
                                                     ~~~~,\nonumber 
\end{eqnarray}
induced by parity conservation.
Because of (\ref{e:vVVA}) and (\ref{e:vprop})   
all the possible components of $\vllp$ can be simply 
derived from the following six components 
\begin{eqnarray}
     \vl0&=&{{\zeta}^2}~\xi^2~~, \nonumber \\
     \vt0&=&\eta+{\frac{1}{2}} \xi~~, \nonumber  \\
     \vtl0&=&{\frac{1}{\sqrt 2}} \zeta\xi {\sqrt {\eta+\xi}}~~,
                                    \label{e:vcomp}  \\
     \vtt0&=&-{\frac{1}{2}} \xi~~, \nonumber \\
     \vht&=& \sqrt {\eta(\eta+\xi)}~~, \nonumber  \\
     \vhtl&=&{\frac{1}{\sqrt 2}} \zeta \xi \sqrt{\eta} 
                  ~~, \nonumber
\end{eqnarray}
where the indices $L,\  T,\  TL$ and $TT$ correspond to 
$(\lambda,\lambda^{\prime}) = (0,0),\ (1,1),\ (1,0)$ and $(1,-1)$; 
$\xi=Q^2/\qlab^2$~~ and  $\eta=\t2g$. 
Note that the definitions (\ref{e:vcomp}) of the $v's$ include the 
appropriate factors of  $\zeta = (\qlab/\qcm)$ which are necessary 
because we calculate the nuclear matrix elements in the c.m. frame. 
\par
The cross section (\ref{e:dsigmaTT}) is the sum of a purely 
electromagnetic term due to the one-photon exchange process and of 
the $\gamma-Z^0$ interference term: 
\begin{equation}
\left(\dsigma\right)=
       \left(\dsigma\right)_{[\gamma]} + 
       \left(\dsigma\right)_{[\gamma-Z^0]}
                                           ~~~~.\label{e:dsigma}
\end{equation}
The dependence of these two terms on the angle  $\phi$ between the 
reaction plane and the scattering plane can be easily separated out
observing that the $T$-transition matrices depend on $\phi$ through a phase 
\begin{equation}
T_{s m_s \lambda m_d}=e^{i (\lambda + m_d) \phi} ~~~\tmatrix
            ~~~~.\label{e:Tt}
\end{equation}
The reduced $t$-matrices so defined depend only on the polar nucleon 
emission angle $\tetacm$ and on the relative momentum 
$|{\bf p}_{c.m.}|$. 
\par
The two cross sections defined in (\ref{e:dsigma}) can be written in the form 
\begin{eqnarray}
          \left(\dsigma\right)_{[\gamma]}& ={\dfrac{\sigma_M}{M_d}}
          \left({\cal F} + h {\cal F}^{(h)} \right) ~~~~,\nonumber \\
          \left(\dsigma\right)_{[\gamma-Z^0]}& ={\dfrac{\sigma_M}{M_d}}
          \left({\cal G} + h {\cal G}^{(h)} \right) ~~~~, 
          \label{e:dsigmaFG}
\end{eqnarray}  
where
\begin{eqnarray}
        {\cal G}       &=& \geff (\gve {\cal G}_1 + \gae {\cal G}_2) 
                                                      ~~~~, \nonumber \\
        {\cal G}^{(h)} &=& \geff (\gae {\cal G}_1 + \gve {\cal G}_2) ~~~~. 
        \label{e:GGh}
\end{eqnarray}
The functions ${\cal F}, {\cal F}^{(h)}, {\cal G}_1, {\cal G}_2$ are given by 
\begin{eqnarray}
       {\cal F} &=& \vl0\fem{L} + \vt0\fem{T} \nonumber \\
          &+& \cc2fi ~\vtt0\fem{TT} + \cfi~ \vtl0\fem{TL} ~~~~,\nonumber \\
  {\cal F}^{(h)} &=& - \sfi ~\vhtl \fhemtl ~~~~,\nonumber \\
      {\cal G}_1 &=& \vl0\femv{L} + \vt0\femv{T} \label{e:FFhG12} \\
          &+& \vtl0 (\cfi~\femv{TL}+\sfi~\fema{TL}) \nonumber \\
          &+& \vtt0 (\cc2fi~\femv{TT}+\ss2fi~\fema{TT}) ~~~~,\nonumber \\
      {\cal G}_2 &=& \vht\fhemat + \vhtl (\cfi~\fhematl+\sfi~\fhemvtl) ~~~~,
              \nonumber
\end{eqnarray} 
in terms of the structure functions
\begin{eqnarray}
       \femllp  &=&  2 \dellp Re(w^{(em)}_{\lambda\lambda^{\prime}}) 
                         ~~~~,\nonumber \\
       \fhemllp &=& -2 \dellp Im(w^{(em)}_{\lambda\lambda^{\prime}}) 
                         ~~~~,\nonumber \\
       \femvllp &=& -2 \dellp Re(w^{(em-V)}_{\lambda\lambda^{\prime}}) 
                         ~~~~,\label{e:femVA} \\
       \fhemvllp&=&  2 \dellp Im(w^{(em-V)}_{\lambda\lambda^{\prime}}) 
                         ~~~~,\nonumber \\
       \femallp &=&  2 \dellp Im(w^{(em-A)}_{\lambda\lambda^{\prime}}) 
                         ~~~~,\nonumber \\
       \fhemallp&=& -2 \dellp Re(w^{(em-A)}_{\lambda\lambda^{\prime}}) 
                         ~~~~, \nonumber
\end{eqnarray}
where
\begin{eqnarray}       
        \wemllp &=& {\frac{1}{3}} \sum_{s m_s} \tem \temsllp ~~~~,\nonumber \\
        \wemvllp&=& {\frac{1}{3}} \sum_{s m_s} \left(\tem \tncvsllp + 
                                 \tncv \temsllp\right) ~~~~,\label{e:wemVA} \\ 
        \wemallp&=& {\frac{1}{3}} \sum_{s m_s} \left(\tem \tncasllp +
                                           \tnca \temsllp\right) ~~~~.
        \nonumber 
\end{eqnarray}
The hadronic tensors $w_{\lambda\lambda^\prime}$ satisfy the symmetry 
relations
\begin{eqnarray}
    w^{*}_{\lambda\lambda^\prime}
              &=& w_{\lambda^\prime\lambda} ~~~~,\label{e:wpropa} \\
    w^{(em)}_{-\lambda-\lambda^\prime}
              &=&(-1)^{\lambda+\lambda^\prime} 
                 w^{(em)}_{\lambda\lambda^\prime} ~~~~,\label{e:wpropb} \\
    w^{(em-V)}_{-\lambda-\lambda^\prime}
              &=&(-1)^{\lambda+\lambda^\prime} 
               w^{(em-V)}_{\lambda\lambda^\prime} ~~~~,\label{e:wpropc} \\
    w^{(em-A)}_{-\lambda-\lambda^\prime}
              &=&(-1)^{1+\lambda+\lambda^\prime} 
               w^{(em-A)}_{\lambda\lambda^\prime} ~~~~,\label{e:wpropd}
\end{eqnarray}
which have already been used together with (\ref{e:vVVA}) and (\ref{e:vprop}) 
to write Eq.(\ref{e:FFhG12}) in terms of 
$\lambda=0,1 ; -\lambda\leq\lambda^\prime\leq\lambda$ only. 
\par
The property (\ref{e:wpropa}) is an immediate consequence 
of definitions (\ref{e:wemVA}). 
The other properties (\ref{e:wpropb} - \ref{e:wpropd}) derive from 
the symmetry relations induced on the $t$-matrix elements by the parity 
conservation:
\begin{eqnarray}  
   t^{(em),(NC)V}_{s -m_s -\lambda -m_d} &=& (-1)^{1+s+m_s+\lambda+m_d} 
                        t^{(em),(NC)V}_{s m_s \lambda m_d} ~~~~, \nonumber \\
                                  &    \cr
   t^{(NC)A}_{s -m_s -\lambda -m_d} &=& (-1)^{s+m_s+\lambda+m_d} 
                         t^{(NC)A}_{s m_s \lambda m_d} ~~~~. \label{e:t}
\end{eqnarray}
\par
The structure function $f^{(em)}_i$ 
are the usual structure functions
of the PC e-d inelastic scattering. 
The functions $f^{(em-V)}_i$,$f^{(em-A)}_i$
are the additional structure functions arising in ewk 
inelastic scattering from the interference between the em current
and the weak vector, axial-vector currents. 
\par
Integrating Eq.(\ref{e:dsigmaFG}) over the outgoing nucleon solid angle, we 
recover the well-known expression of the inclusive cross section, first 
given by Walecka ~\cite{Walecka75} on the basis of symmetry considerations 
and covariance requirement. In such integration, all the TL and TT 
interference terms drop to zero and the surviving 5 exclusive structure 
functions transform into the inclusive response functions.
\subsection{Nucleon electromagnetic and weak form factors}
The general 
expressions of the matrix elements of the single-nucleon ewk currents 
consistent with Lorentz covariance and with parity and time-reversal 
invariance are 
\begin{eqnarray}
  \Jem_\mu  &=& \dfrac{1}{2M(1+\tau)} \bar u(p^\prime) 
                \Big[ G_E (p+p^\prime)_\mu 
                    + G_M (2M \tau \gamma_\mu 
                           +i \sigma_{\mu\nu} q^\nu) \Big] u(p) 
                                                      ~~~~,\label{e:Jem} \\ 
  \Jncv_\mu &=& \dfrac{1}{2M(1+\tau)} \bar u(p^\prime) 
                \Big[ {\tilde G_E} (p+p^\prime)_\mu 
                    + {\tilde G_M} (2M \tau \gamma_\mu 
                                   +i \sigma_{\mu\nu} q^\nu) \Big] u(p) 
                                                      ~~~~,\label{e:Jncv} \\ 
  \Jnca_\mu &=& \bar u(p^\prime) \big[\tilde G_A \gamma_\mu  
            + i(\tilde G_P/M) q_\mu \big] \gamma_5 u(p) ~~~~,\label{e:Jnca} 
\end{eqnarray}
where $M$ is the nucleon mass, $\tau = Q^2/4M^2$, 
$p$ and $p^\prime$ are the four-momentum of the incoming and outgoing nucleon, 
respectively. 
\par
In the following, we do not need to care about the induced pseudoscalar 
current because it does not 
contribute to observables in PV electron scattering to leading order in 
ewk coupling. 
\par
We have chosen the Sachs form of $\Jem$ and $\Jncv$ because the study of 
the PC deuteron electrodisintegration has revealed that, unlike the Dirac 
form of $\Jem$, it leads to non relativistic (NR) results close to the 
full theory results, minimizing the effect of the relativistic corrections. 
From the same analysis we also know that the cross section is almost 
insensitive to meson exchange and isobar excitation currents in the 
QE region. In conclusion we shall not consider relativistic corrections 
and interaction currents in our calculations.
\par
From the structure of the em and weak-vector current operators in terms 
of the SU(3)-singlet and -octet currents it follows that the nucleon 
weak-vector form factors are given by
\begin{eqnarray}
  {\tilde G}_{E,M}(Q^2) &=& \dfrac{1}{2} \xi_V^{T=1} G_{E,M}^V(Q^2) \tau_3 
                  + \dfrac{\sqrt{3}}{2} \xi_V^{T=0} G_{E,M}^S(Q^2) 
                                        \nonumber \\
                 &+&                    \xi_V^{(0)} G_{E,M}^{(s)}(Q^2) 
                 ~~~~,\label{e:GEMW}
\end{eqnarray}
with  $\tau_3$=+1,~-1 for the proton and neutron, respectively. 
~$G_{E,M}^{S(V)}$ is the isoscalar (isovector) combination of the em 
Sachs form factors, $G^{(s)}_{E,M}$ is the strange-quark contribution and 
the couplings are appropriate linear combinations of quark weak-vector 
charges. In the Standard model they have the values
\begin{equation}
 \xi^{T=1}_V = 2(1-2 sin^2 \vartheta_W) ~~~~,~~~~
 \sqrt{3} \xi^{T=0}_V = -4 sin^2 \vartheta_W ~~~~,~~~~ 
 \xi^{(0)}_V = - 1
                  ~~~~~.\label{e:csivector}
\end{equation}
\par
According to ~\cite{MusolfPR94}, we take
the strangeness weak vector form factors in the form
\begin{eqnarray}
  \GES &=&  \rhos \tau G_D^V(Q^2) (1+\laes \tau)^{-1}
                                               ~~~~~, \nonumber \\
                 & &                                  \label{e:GEMstrani} \\
  \GMS &=&   \mus  G_D^V(Q^2) (1+\lams \tau)^{-1}
                                               ~~~~~, \nonumber 
\end{eqnarray}
which is an extension of the Galster parametrization ~\cite{Galster71} 
commonly used for the nucleon em form factors
\begin{eqnarray}
   G_E^p(Q^2) &=& G_D^V(Q^2)~~~~~,~~~~~
   G_E^n(Q^2) = -\mu_n \tau G_D^V(Q^2) (1+5.6 \tau)^{-1} ~~~~~\nonumber \\
         & &                       \label{e:Galster} \\
   G_M^p(Q^2) &=&  \mu_p  G_D^V(Q^2)~~~~~,~~~~~
   G_M^n(Q^2) =  \mu_n  G_D^V(Q^2)  ~~~~~,\nonumber
\end{eqnarray}
where ~$G_D^V(Q^2) = (1+Q^2/M_V^2)^{-2}$ , 
with a cut-off mass squared $M_V^2$=0.71 $GeV^2$. 
\par
Expression ~(\ref{e:GEMstrani}) of $G^{(s)}_E$ implements the only 
theoretical constraint about the strangeness form factors. The nucleon 
has no net strangeness, so that $G^{(s)}_E(0)=0$  
and the low $Q^2$ behaviour of $G^{(s)}_E$ 
is characterized by the dimensionless strangeness radius
\begin{equation}
  \rhos \equiv \left [ \dfrac{d~G_E^{(s)}(Q^2)}{d~\tau} \right ]_{\tau=0} 
                         ~~~~~~. \label{e:rhostrano} 
\end{equation}
Also commonly used in the literature is the Dirac strangeness radius 
\begin{equation}
  \rdues \equiv - 6 \left [ \dfrac{d~F_1^{(s)}(Q^2)}{d~Q^2} \right ]_{Q^2=0} 
                                      ~~~~~~~~~.\label{e:r2s}
\end{equation}
\par
Because of the well known relations between the Sachs and the Dirac form 
factors, $\rhos$, $\mus$ and $\rdues$ are linearly related by 
\begin{equation}
    \rhos = - \dfrac{2}{3} M^2 \rdues  - \mus    
                                     ~~~~~~~~~.\label{e:r2srhomu}
\end{equation}
\par
Very little is known about the values of $\mus$ and $\rdues$ even if many 
calculations of the strangeness vector form factors have been carried out 
using different approaches (lattice calculations, effective Lagrangian, 
dispersion relations, hadronic models). 
The predictions of the strangeness moments are quite different in different 
approaches and can also largely vary within a given approach because of the 
need of additional assumptions and approximations. 
In particular, $\rdues$ is predicted to be positive in the dispersion 
theory analysis of the nucleon isoscalar form factors 
~\cite{Jaffe89,Hammer96}, 
of the same order of magnitude but negative by the 
chiral quark-soliton model ~\cite{Kim95} and negative but of 
two order of magnitude smaller by the kaon-loop 
calculations ~\cite{MusolfZfP94}. 
A negative value of $\rdues$ is also preferred by the analysis 
~\cite{Garvey93} of the $\nu p/{\bar \nu} p$ elastic scattering data 
~\cite{Ahrens87}  which, however, has been criticized for the use 
of a unique cut-off mass for the three SU(3) axial-vector form factors. 
\par 
The different existing models widely disagree also about sign and magnitude 
of $\mus$ which is predicted to range from 
$\mus=0.4 \pm 0.3 ~\mu_N$ ~\cite{HongPark93} in the chiral hyperbag model to 
$\mus=-0.75 \pm 0.30 ~\mu_N$ ~\cite{Leinweber96} using QCD equalities 
among the octet baryon magnetic moments. 
\par
Clearly, a model independent determination of the strangeness moments 
and, possibly, of the $Q^2$-dependence of the strangeness form factors, 
can only come from the experiments. 
\par
Analogously to (\ref{e:GEMW}), the axial-vector form factor can be decomposed 
in terms of the 3rd and 8th SU(3) octet components and of the possible 
strange component 
\begin{equation}
  {\tilde G}_A(Q^2) = \dfrac{1}{2} \xi_A^{T=1} G_A^{(3)}(Q^2) \tau_3 
                 + \dfrac{1}{2} \xi_A^{T=0} G_A^{(8)}(Q^2) 
                 + \xi_A^{(0)} G_A^{(s)}(Q^2) 
                 ~~~~, \label{e:GAP} 
\end{equation}
with coupling constants dictated at the tree level by the quark axial charges 
\begin{equation}
 \xi^{T=1}_A = - 2 ~~~~,~~~~ 
 \xi^{T=0}_A = 0 ~~~~,~~~~ 
 \xi^{(0)}_A = 1
                  ~~~~~.\label{e:csiaxial}
\end{equation}
\par
Note that in this limit the isoscalar component of ${\tilde G}_A$ fully comes 
from the strange quark contribution. Information on the $Q^2=0$ value of the 
SU(3) octet form factors derives from charged current weak interactions. 
From neutron $\beta$-decay and strong isospin symmetry it follows 
$G^{(3)}_A(0) = (D+F) \equiv g_A = 1.2601 \pm 0.0025$ ~\cite{PDG96}, 
while from hyperon $\beta$-decays and flavour SU(3) symmetry it follows 
$G^{(8)}_A(0) = (1/\sqrt{3}) (3F-D) = 0.334 \pm 0.014$ ~\cite{Close93} , 
%
%
$D$ and $F$ being the 
associated SU(3) reduced matrix elements. 
The $Q^2$ dependence of these form factors can be adequately parametrized 
with a dipole form 
\begin{equation}
    G_D^A(Q^2) = (1+Q^2/M_A^2)^{-2}  ~~~~~, \label{e:GAdip}
\end{equation}
with a cut-off mass $M_A$=1.032 GeV. 
The same dipole form is suggested in \cite{MusolfPR94}
for the strange axial vector form factor 
\begin{equation}
  \GAS =  \etas g_A G_D^A(Q^2) (1+\lambda_A^{(s)} \tau)^{-1}
                                 ~~~~~.  \label{e:GAstrano}
\end{equation}
Here again, lacking theoretical constraints on $G^{(s)}_A(0)$ and 
because of the model dependence of the theoretical estimates, 
values of $\etas=G_A^{(s)}(0)/g_A$  have to be extracted from the 
experiments. As mentioned in the Introduction, the first indications 
came from the BNL $\nu p / {\bar \nu} p$ experiment and from the EMC data.
\par
As for the weak coupling constants we emphasize that the values 
(\ref{e:csivector}) and (\ref{e:csiaxial}) are those predicted by 
the ewk standard model at the tree-level.
In a realistic evaluation of the amplitude of any electron-hadron process 
one has to consider the radiative corrections to these values. 
Such corrections $R^{(a)}_{V,A}$, 
amounting to a factor (1+$R^{(a)}_{V,A}$) in all the coupling 
constants except in $\xi^{T=0}_A$ which becomes $\sqrt{3} R^{T=0}_A$, 
are very difficult to calculate because they receive 
contributions from a variety of processes (higher-order terms in ewk 
theory, hadronic physics effects,...).  They have been estimated by 
various authors (for a review see Ref.\cite{MusolfPR94}~ and citations therein) 
using different approaches and approximations with results in qualitative 
agreement. 
More precisely, $R^{(a)}_V$ are estimated to be of the order of a few percent 
and $R^{(a)}_A$ of the order of some tenth of percent. Therefore, while 
$R^{(a)}_V$ can be neglected, the radiative corrections $R^{(a)}_A$ must, 
in principle, be taken into account. 
\subsection{Asymmetry}
As said in the Introduction we are interested in the helicity asymmetry 
of the coincidence cross section, which is defined as
\begin{equation}
  {\cal A}(\tetacm,\phi) = \frac{\sigma(h=+1) - \sigma(h=-1)} 
                                    {\sigma(h=+1) + \sigma(h=-1)} 
                             ~~~~~,\label{e:Adef}
\end{equation}
where $\sigma(h=\pm 1)$ is the exclusive cross section for electrons 
polarized parallel $(h=+1)$ and antiparallel $(h=-1)$ to their momenta. 
From Eq.(\ref{e:dsigmaFG}) we have
\begin{equation}
  {\cal A}(\tetacm,\phi) = \frac{{\cal F}^{(h)} + {\cal{G}}^{(h)}} 
                             {{\cal F} + {\cal{G}}} 
                      \simeq \frac{{\cal F}^{(h)} + {\cal G}^{(h)}}{\cal F} 
                             ~~~~~,\label{e:AFG}
\end{equation}
because ${\cal G}$ is negligible with respect to ${\cal F}$. 
The term ${\cal F}^{(h)}$ is the purely em contribution to the helicity 
dependent part of the cross section (proportional to the fifth 
structure function $f^{(em)h}_{TL}$) 
which vanishes in coplanar geometry (see Eq.(\ref{e:FFhG12}) ). 
Thus, considering the in-plane kinematics and to leading order in G, 
the helicity asymmetry is given by the interference of weak and em amplitudes 
and reads 
\begin{eqnarray}
    {\cal F ~A}(\tetacm) 
      &=& ~\geff \Big[\gve (\vht f^{(em-A)h}_T 
                 \pm \vhtl f^{(em-A)h}_{TL})    \label{e:sigma0A} \\
      &+& ~\gae (\vl0 f^{(em-V)}_L + \vt0 f^{(em-V)}_T \nonumber \\
      &\pm& \vtl0 f^{(em-V)}_{TL} + \vtt0 f^{(em-V)}_{TT}) \Big] 
                                   ~~~~, \nonumber
\end{eqnarray}
where the sign $\pm$ corresponds to $\phi = 0^\circ , 180^\circ$. 
Note that the z-axis is along {\bf q} and the y-axis is normal to the 
reaction plane in the direction ${\bf k}_e \times {\bf k}^\prime_e$. 
\par
In an experiment, the finite angular acceptance of the spectrometers 
makes it unavoidable to also collect nucleons emitted out-of-plane and thus, 
apparently, to include in the measured asymmetry the effect of the PC 
contribution ${\cal F}^{(h)}$ which could mask the PV asymmetry. 
Actually, the experimental results 
correspond to an average of the theoretical 
expression (\ref{e:AFG}) over the spectrometer solid angle. 
In such an average the influence of the 
fifth structure function should vanish because $f^{(em)h}_{TL}$ enters 
the cross section multiplied by a factor sin$\phi$, if the spectrometer 
is exactly centered and symmetrical. 
Since this is not the case in a real experiment, one has to consider 
the PC asymmetry in the out-of-plane kinematics close to the electron 
scattering plane. In the next section we shall give a quantitative 
estimate of how symmetrical the hadron spectrometer must be in order to 
make possible to extract the PV asymmetry from the measured asymmetry. 
\par
The PV exclusive asymmetry (\ref{e:sigma0A}) shows a very rich structure. 
In fact, it depends on six structure functions which probe different 
components of the weak vector and axial currents. In principle, the effects 
of a particular component could be singled out. For example, the longitudinal 
parts of the weak currents appearing in longitudinal-transverse structure 
functions $f^{(em-A)h}_{TL}$ and  $f^{(em-V)}_{TL}$ could be derived from 
the difference of $\Ateta$ measured at the same $\tetacm$ 
in the half-plane $\phi = 0^\circ$ and $\phi = 180^\circ$. 
Other structure functions could be isolated by some generalized Rosenbluth 
decomposition. However, it does not seem to us sensible to further 
elaborate on this point since such a program is, at the moment, completely 
beyond experimental feasibility.
\par
To get an idea of how the exclusive asymmetry depends on the weak form factors
of the nucleon it is convenient to consider the simplified form of
${\cal A}(\tetacm=0^\circ)$ obtained in the PWIA model, 
which consists in taking into account only the dominant contribution arising 
from the knocked-out nucleon in plane wave (PW) approximation for the 
final states and in S-wave deuteron state. 
In that approximation, which accurately reproduces
the full theory results for nucleons detected in forward direction
($\tetacm \simeq 0^\circ$), one finds 
\begin{equation}
   {\cal A}(\tetacm = 0^\circ) \simeq 
          - 2 ~\geff~ {\frac{2~ \gve \vht \sqrt{\tau}  G_M \tilde G_A 
          + \gae (\vl0 G_E \tilde G_E + 2 \vt0 \tau G_M \tilde G_M)}
            {\vl0 G_E^2 + 2 \vt0 \tau G_M^2}}
       ~~~. \label{e:Azero}
\end{equation}
\par     
Therefore, a measurement of the asymmetry for neutrons emitted at 
$\tetan=0^\circ$, or, equivalently, for protons outgoing at 
$\tetacm=180^\circ$, allows one to determine the neutron weak form 
factors. The difference with the asymmetry in the ${\vec e}d$ inclusive 
reaction can be easily appreciated recalling the approximate form of 
the inclusive asymmetry (the so-called static approximation~\cite{MusolfPR94}) 
which is similar to expression (\ref{e:Azero}) but depends on the 
incoherent sum of the contributions coming from proton and neutron. 
Thus, while the inclusive asymmetry is sensitive to the average 
of the nucleon form factors, the exclusive asymmetry feels the 
influence of the individual form factors. This enhanced sensitivity  
might make it interesting to measure the exclusive asymmetry 
not withstanding the reduced rate of the coincidence cross section. 
\par
Apart from minor differences deriving from the not-completely-covariant 
treatment of the ${\vec e}d$ inelastic scattering, expression (\ref{e:Azero}) 
coincides with the PV electron-free nucleon asymmetry. 
\par
Thus, the limiting cases well known from the analysis of the PV ${\vec e}p$ 
scattering apply in the PV ${\vec e}d$ exclusive disintegration, namely 
the magnetic interactions dominate for electron backwardly scattered, while 
the electric interactions play a major role for electron forwardly scattered. 
The effect of the axial form factor is suppressed because of the small 
value of the electron weak vector charge, as is already clear in the general 
expression (\ref{e:sigma0A}). 
\section{Results}
In this paper we limit our considerations to the low momentum transfer 
region ($Q^2 \simeq 0.1 (GeV/c)^2$ as in SAMPLE experiment) in order to 
minimize the impact of the uncertainties in the $Q^2$ dependence of the 
strange form factors. 
To be explicit, in the calculations we take the 
values $\laes=5.6,~~\lams=0,~~\laas=0$ of the parameters which determine 
the $Q^2$ fall-off of the expressions (\ref{e:GEMstrani},\ref{e:GAstrano}) 
of the strange form factors. Because of the low $Q^2$ 
considered, these assumptions should not be crucial. 
\par
Furthermore, as reference values for the axial radiative corrections we 
adopt $\RATZERO=-0.62$ and $\RAT1=-0.34$ given by Musolf and Holstein 
\cite{MusolfPLB90} using for the hadronic contributions the so-called 
best estimates for the weak meson-nucleon vertices of 
Ref.\cite{Desplanques80}. Finally, we use the value $\etas=-0.12$, 
deduced from the neutrino scattering experiment, of the strangeness 
axial charge and, lacking a reliable estimate of the 
radiative correction to the strangeness axial coupling constant, 
we take $\RAS=0$. 
All the constants entering the calculations, except the 
strangeness radius and magnetic moment, having been fixed, we can concentrate 
on the effect of $\rdues$ and $\mus$ on the exclusive asymmetry. 
\par
In the following we report the proton asymmetry as a function of the proton 
polar angles $\tetacm$. To distinguish the half-plane 
$\phi=0^\circ$ and $\phi=180^\circ$, we assign the positive (negative) 
sign to $\tetacm$ for protons emitted in 
the half-plane $\phi=0^\circ$ ($\phi=180^\circ$). 
The same figures can be used to deduce the neutron asymmetry ${\cal A}_n$, 
i.e. the PV asymmetry in the (${\vec e},e^\prime n$) reaction. 
Obviously, the value of ${\cal A}_n$ for neutrons outgoing 
at ($\tetacm,\phi$) corresponds to the value of the proton asymmetry 
at ($\pi - \tetacm , \pi + \phi$). 
\par
Let us start considering the angular distribution of $\Apteta$ in the QE 
region for backward scattering electron ($\vartheta_e=160^\circ$) 
where the role of $G_E^{(s)}$ is strongly suppressed. 
\par
In Fig.1 we plot the c.m. angular distribution of $\Apteta$ calculated 
with Jaffe's values \cite{Jaffe89} of the strangeness radius 
($\rdues$=0.16~fm$^2$) and magnetic moment ($\mus=-0.31~\mu_N$). 
In order to study the dependence of the asymmetry on the NN potential 
models, we have used 
the deuteron wave functions as well as np continuum wave functions 
calculated with the Paris potential \cite{PARIS80} and with 
the folded diagram potential OBEPF \cite{OBEPF92} 
which gives predictions of the NN data in close agreement with the 
full Bonn potential.
Actually, the final state interactions are taken into account 
in the multipole amplitudes up to L=6, 
while all the other multipole amplitudes are evaluated in free-wave 
approximation, as described in Ref.\cite{MR90}. 
The angular distribution of $\Apteta$ is characterized by two minima 
(note that $\Apteta$ is negative in all the range of $\tetacm$) almost 
symmetric with respect to {\bf q} and by a maximum at $180^\circ$, where 
the asymmetry is a factor 1.5 higher than at $\tetacm=0^\circ$. 
Obviously such a maximum at backward proton angles corresponds to 
the emission of neutrons at forward angles. 
\par
The very weak dependence on the NN potential model in all the angular range 
suggests some reason beyond the fact that the asymmetry is defined 
as a ratio of cross sections, which could be the 
dominance of the transitions from the S-wave deuteron state. 
\begin{figure}
\centerline{\psfig{figure=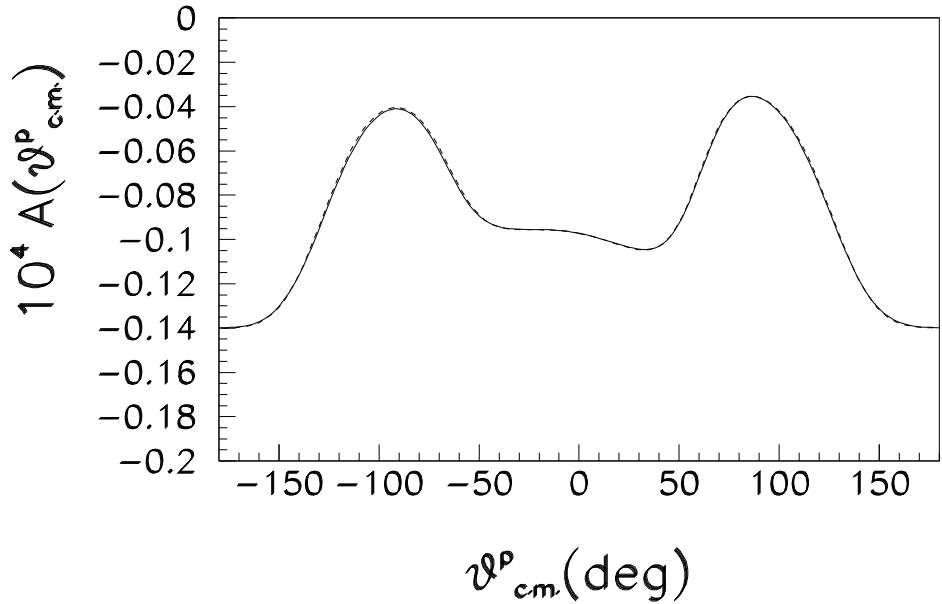,width=14.0cm}} 
\caption{
   Angular distribution of the proton asymmetry ${\cal A}(\tetacm^p)$ 
   in the quasi-elastic region at 
   $Q^2$=0.1$(GeV/c)^2$, $\vartheta_e=160^\circ$, 
   with $\mus=-0.31$ , $\rdues=0.16$ ~\cite{Jaffe89}. 
   Calculations are with the Paris potential (full line) and the 
   OBEPF potential (dashed line). 
}
\end{figure}
\par
The advantage of the exclusive deuteron ewk disintegration which we 
already have alluded to lies in the possibility of performing 
simultaneous measurements of $\Ap$ at $\tetacm = 0^\circ$ and at 
$\tetacm = 180^\circ$ or equivalently of $\An$ at $\tetan = 0^\circ$. 
When combined with the results of the ${\vec e}p$ asymmetry they can 
lead to an accurate determination of $\mus$. The comparison of $\Ap$ at 
$\tetacm = 0^\circ$ with the asymmetry in the ${\vec e}p$ scattering 
should serve as a check of the exclusive experiment. 
\begin{figure}
\centerline{\psfig{figure=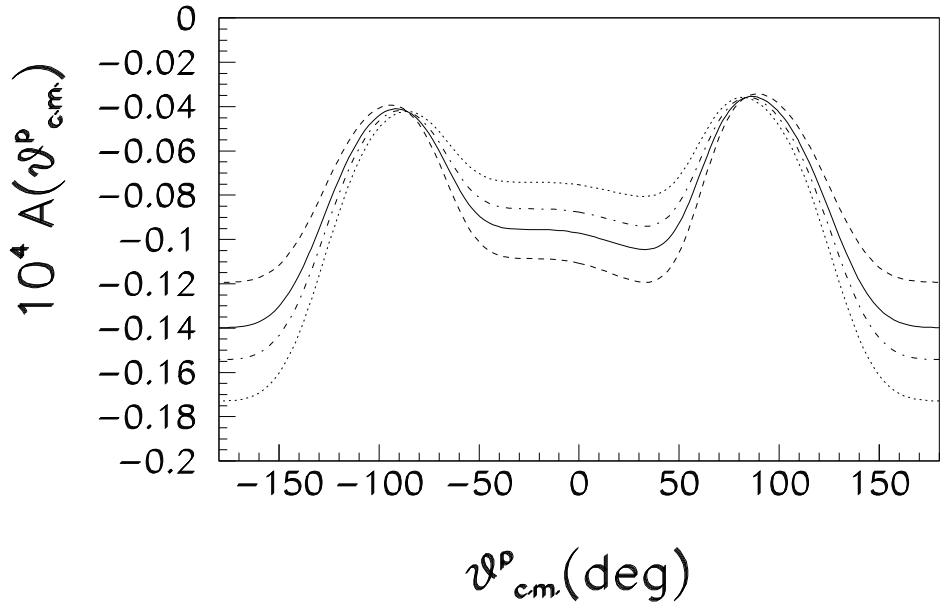, width=14.0cm}}
\caption{
  Dependence of the proton asymmetry ${\cal A}(\tetacm^p)$ 
  on the strange magnetic 
  moment $\mus$ in the case of $\vartheta_e=160^\circ$. 
  The solid line is the same as in Fig.1. 
  The other curves are for $\mus$=-0.75 ~\cite{Leinweber96} 
  (dashed line), $\mus$=0.40 ~\cite{HongPark93} (dotted line) 
  and for $\mus$=0 (dot-dashed line). 
}
\end{figure}
\par
Intuitively, the exclusive cross section $(e,e^\prime N)$ at the QE peak 
($\tetacm^N = 0^\circ$), where the detected nucleon is ejected 
in the direction of ${\bf q}$, should be very close to the cross section for 
electron scattering on free nucleon. This is confirmed by actual calculations 
which give $\Ap = - 0.096~10^{-4}$ and $\An = - 0.140~10^{-4}$ 
with the same choice 
of form factors and in the same kinematical conditions as in Fig.1. 
For comparison, the corresponding values in the ${\vec e}d$ exclusive 
asymmetry are $-0.097~10^{-4}$ and $- 0.140~10^{-4}$, respectively. 
This fact will be exploited later on in the discussion of the precision 
reachable in the determination of $\mus$. 
\par
The knowledge of $\An$ can be exploited directly and through the ratio 
$\Ap/\An$ where the systematic uncertainties cancel to a very large extent 
since $\Ap$ and $\An$ have been measured under 
exactly the same experimental conditions. 
A similar cancellation of the systematic errors has been envisaged by the 
SAMPLE experiment which intends to use the ratio $\Ap/\Ad$, where $\Ad$ is 
the asymmetry in the inclusive ${\vec e}d$ inelastic scattering. 
\par
To show the effect on the asymmetry of variations in the strangeness 
magnetic moment we report in Fig.2 our results for the Paris 
potential and for a selected set of 
predictions of $\mus$. Among the values given by the different models 
we have chosen those defining the theoretical range of $\mus$, 
i.e. $\mus=-0.75~\mu_N$ \cite{Leinweber96} and 
$\mus=0.4~\mu_N$ \cite{HongPark93}. 
Also reported are the curves corresponding to Jaffe's value of $\mus$ 
\cite{Jaffe89} and to $\mus=0$. 
Note that the Dirac strangeness radius has been held fixed, 
$\rdues=0.16$~fm$^2$ as deduced by Jaffe. 
This comparison makes evident the strong sensitivity on $\mus$ of the 
asymmetry for electrons scattered in the backward direction. 
\par
To be more quantitative on the precision reachable in a determination 
of $\mus$, let us consider again the PW expression 
(\ref{e:Azero}) of the proton and neutron asymmetry at 
$\tetacm^N=0^\circ ~~(N=p,n)$ and write it in the form 
\begin{equation}
 {\cal A}_N \equiv {\cal A}_N^0 
            \Big(1 + a_N ~\rhos + b_N ~\mus 
                   + c_N ~\RAT1 + d_N ~\RATZERO + e_N ~\etas \Big) 
                      ~~~~,\label{e:abcd}
\end{equation}
which exibits the dependence on the unknown strangeness radius, 
magnetic moment and axial charge (in unit of $g_A$) and 
on the radiative corrections to the axial-vector coupling constant. 
Actually, the possible modification due to 
radiative corrections $R^{(0)}_A$ of the strange-quark axial coupling 
constant is understood in the last term in (\ref{e:abcd}). 
\begin{table}[h]
\begin{tabular}{|c|c|c|c|c|c|c|}      
\hline
           &  &  &  &  &  &    \\
           &${\cal A}^0$  &a  &b  &c  &d  &e     \\
           &  &  &  &  &  &    \\
\hline 
           &  &  &  &  &  &   \\
 proton &$-0.88 \times 10^{-5}$ &$-0.16 \times 10^{-2}$ 
                                &-0.342 &0.256 &-0.072 &-0.256 \\
            &  &  &  &  & &   \\
\hline
           &  &  &  &  &  &    \\
 neutron &$-0.17 \times 10^{-4}$ &$-0.85 \times 10^{-4}$
                                 &0.270 &0.202 &0.057 &0.202 \\
           &  &  &  &  &  &    \\
\hline
\end{tabular}
\caption{Values of the constants entering the expression (\ref{e:abcd}) of 
the proton and neutron asymmetries for $Q^2=0.1 (GeV/c)^2$ and 
$\vartheta_e=160^\circ$. }
\end{table}
\par
The values of ${\cal A}^0_N$ and of the other constants are given in Table~1. 
Note, first of all, the smallness of $a_N$ which fully justifies our previous 
statement about the substantial independence of $G_E^{(s)}$. Second, also 
the influence of $\RATZERO$ is greatly reduced. Finally, we note that 
$c_N$ and $e_N$ have the same value and the same sign 
in the case of the neutron but opposite sign in the case of the proton, 
as a consequence of our choice $\laas = 0$. 
In fact, in this case we have $e_N = - \tau_3 c_N$. 
In conclusion, if we further assume 
that the axial-vector strangeness form factor is known from 
neutrino scattering experiments \cite{LSND89}, 
the precision $\delta\mus$ which can be reached in the determination of $\mus$ 
depends on the experimental accuracy in the measurements and on the 
uncertainty on the isovector axial coupling constant. 
\begin{figure}
\centerline{\psfig{figure=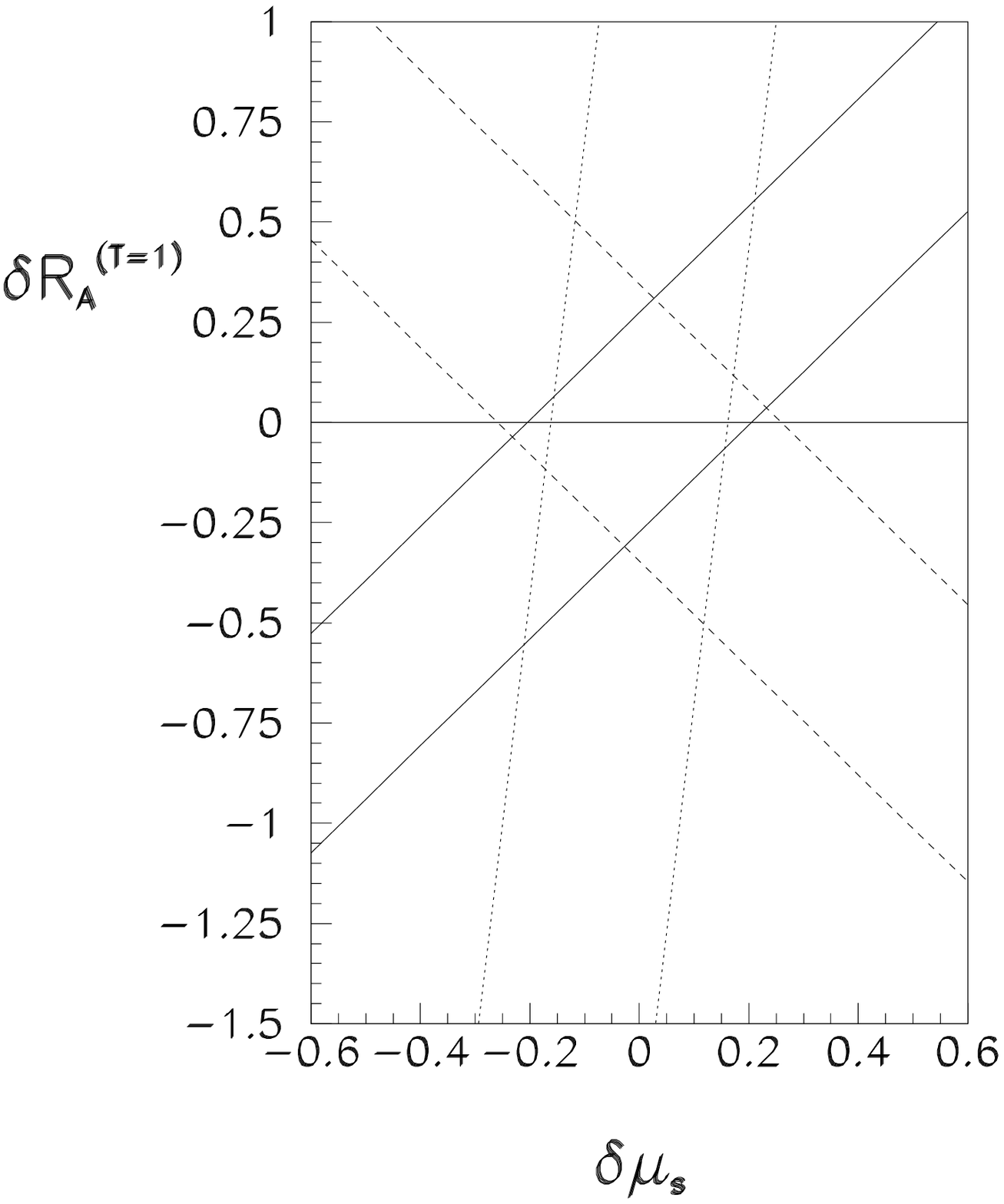, width=10.0cm}} 
\caption{($\RAT1-\mus$) correlation assuming an experimental error
  $\delta{\cal A}/{\cal A}=\pm$7\%. The three error bands are for the 
  results on $\Ap$ (full lines), on $\An$ (dashed lines) and on the 
  ratio $\Ap/\An$ (dotted lines).}
\end{figure}
\par
Then, from Eq.(\ref{e:abcd}) we have that the uncertainty in $\mus$ 
together with the error in $\RAT1$ induce a fractional change in the 
backward-angle asymmetry given by
\begin{equation}
 \dfrac{\delta{\cal A}_{p,n}}{{\cal A}_{p,n}} \simeq
               \Big(b_{p,n} ~\delta\mus + c_{p,n} ~\delta\RAT1 \Big) 
                      ~~~~.\label{e:dAsuA}
\end{equation}
\par
The $\RAT1-\mus$ correlation is 
displayed in Fig.3 where we have assumed an experimental error 
$\delta{\cal A}/{\cal A} = \pm 7 \%$ as in the SAMPLE experiment. 
The three error bands are for the results on $\Ap$ (full lines), 
on $\An$ (dashed lines) and for the ratio $\Ap/\An$ (dotted lines). 
The figure clearly shows that the ratio is almost independent of $\RAT1$ 
and this happens because it enters in the proton and neutron asymmetry 
with the same sign and almost with the same value. 
Clearly, such an experiment allows one to tightly limit the value 
of $\mus$ and of $\RAT1$. 
\begin{figure}
\centerline{\psfig{figure=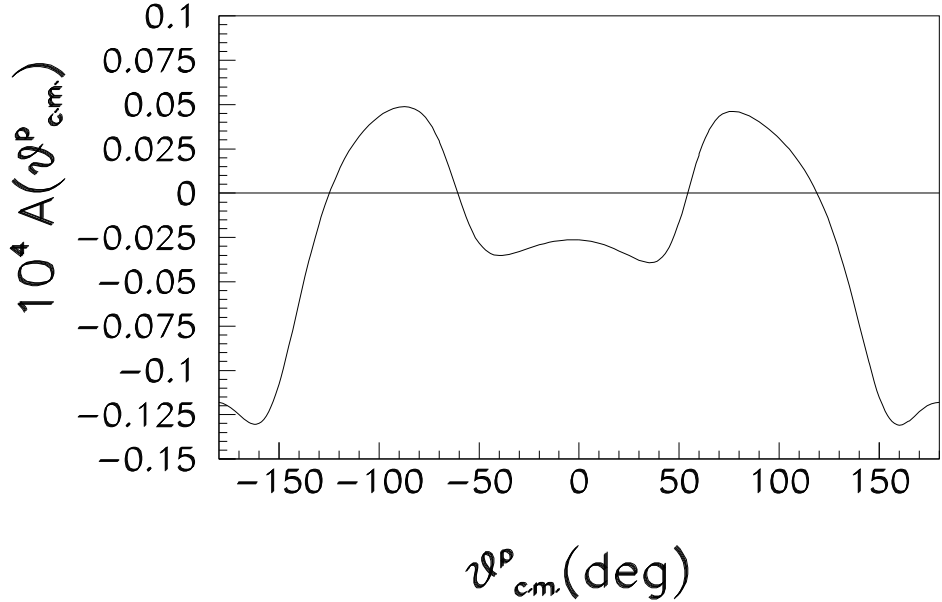, width=14.0cm}}
\caption{The same as in Fig.1 except for $\vartheta_e=15^\circ$.
         Calculations are with the Paris potential.} 
\end{figure}
\par
Conversely, once one has determined $\mus$ and $\RAT1$, 
the ratio $\Ap/\An$ could be exploited for getting information 
on the isoscalar part $\GAT0$ of the axial-vector form factor. 
In fact, contrary to the isovector part $\GATUNO$, 
$\GAT0$ contributes with opposite sign to the proton and neutron asymmetries. 
Further, if the isoscalar coupling constant $\xi_A^{T=0}$ is assumed 
to vanish as predicted by the standard model at the tree-level so that 
$\GAT0$ reduces to the strangeness contribution $\GAS$, 
the effect of the radiative corrections to the strangeness axial-vector 
coupling constant could be studied. 
\par
The exclusive asymmetry $\Apteta$ is plotted in Fig.4 in the same kinematical 
conditions as before, except for the electron scattering angle, 
$\vartheta_e=15^\circ$. In this kinematics, while the 
effects of the axial current are strongly suppressed because 
$\vht \rightarrow 0$, those of the electric weak vector current are 
enhanced ($\vl0/2\vt0 \rightarrow 1$). 
The asymmetry is calculated with Jaffe's values of $\mus$ and $\rdues$ 
\cite{Jaffe89}. Because of its substantial independence of the NN potential 
models, as seen in Fig.1, only the results obtained with the Paris 
potential are drawn.
\begin{figure}
\centerline{\psfig{figure=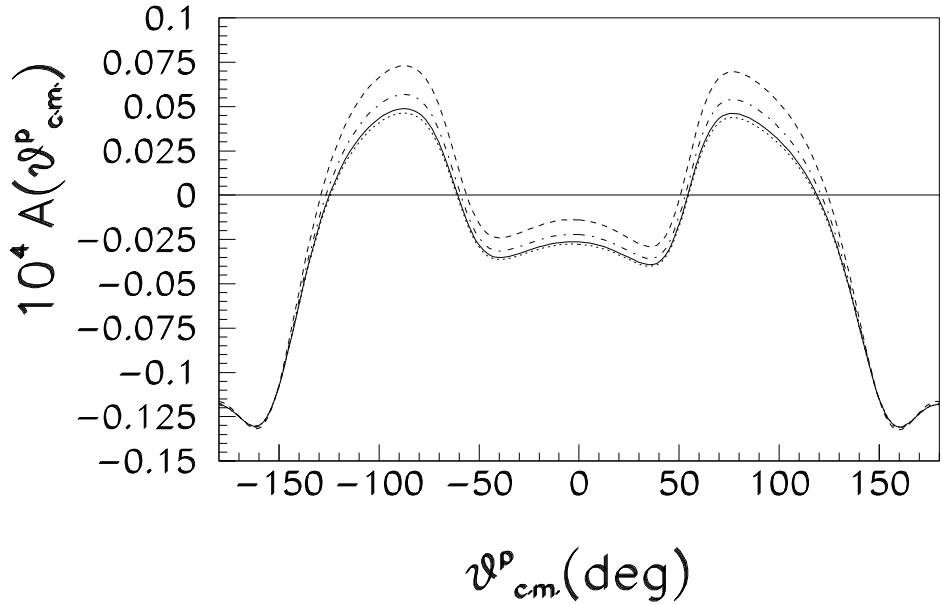, width=14.0cm}} 
\caption{
Dependence of the proton asymmetry ${\cal A}(\tetacm^p)$ 
  on the Dirac strangeness radius $\rdues$ 
  in the case of $\vartheta_e=15^\circ$. 
  The solid line is the same as in Fig.4.  
  The other curves are for $\rdues$=-0.32 ~\cite{Kim95} 
  (dashed line), for $\rdues$=0.21 ~\cite{Hammer96} (dotted line) 
  and for $\rdues$=0 (dot-dashed line). 
}
\end{figure}
\par
The sensitivity to the strangeness radius can be appreciated from Fig.5, 
where we compare our results of $\Apteta$ for a restricted selection 
of predicted $\rdues$, all other parameters being the same. Besides that 
given by Jaffe and $\rdues=0$, we have used the two almost opposite 
values $\rdues=0.21$~fm$^2$, deduced by Hammer et al. \cite{Hammer96} in 
their revised dispersion analysis and $\rdues=-0.32$~fm$^2$ obtained 
by Kim et al. \cite{Kim95} (chiral-quark soliton model). 
At first sight, a measurement of $\Apteta$ in the forward direction or, 
better, at $\tetacm \sim 70^\circ-80^\circ$ where the asymmetry has 
a maximum, 
could lead to discriminate between the different models. 
There is not such sensitivity in the 
asymmetry for neutrons detected at $\tetan=0^\circ$ where the asymmetry 
is a factor 5 higher. The reason is that 
${\tilde G}^n_E = 0.092 G^n_E - G^p_E - G^{(s)}_E$
and $G^p_E$ is so large that variations in $G^{(s)}_E$ cannot be of 
any importance at low $Q^2$. 
\par
Actually, the precision on the extraction of $\rdues$ from such experiment 
is strongly limited by the error induced by the uncertainty in the other 
quantities determining $\Apteta$ and particularly in $\mus$. In fact, the 
same considerations of Ref.\cite{MusolfNPA92} valid for the ${\vec e}p$ 
cross section asymmetry apply to the exclusive ${\vec e}d$ 
cross section asymmetry. This can be seen looking at the values of the 
parameters $a_N, b_N$ reported in Table~2. Clearly, the impact 
of the uncertainty $\delta\mus$ on $\delta\rhos$ is weighted by a large 
factor,in fact $b_p/a_p \sim 3.36$, and even worse in the neutron case 
where $b_n/a_n \sim -49.2$. Also the uncertainty in $G_M^n$ and $G_E^n$ 
can introduce sizeable errors in $\GES$. As pointed out 
in \cite{Hadjimichael92}, 
the asymmetry of the inclusive ${\vec e}d$ cross section seems more 
promising for a determination of $\rdues$ because the influence of $\mus$ 
is suppressed due to the coherent sum of the proton and of the neutron 
contributions.
\begin{table}[h]
\begin{tabular}{|c|c|c|c|c|c|c|}      
\hline
           &  &  &  &  &  &    \\
           &${\cal A}^0$  &a  &b  &c  &d   &e    \\
           &  &  &  &  &  &    \\
\hline 
           &  &  &  &  &  &    \\
 proton &$-0.17 \times 10^{-5}$ &$-0.102$ 
                                &-0.343 &0.067 &-0.019 &-0.067 \\
            &  &  &  &  &  &    \\
\hline
           &  &  &  &  &  &    \\
 neutron &$-0.11 \times 10^{-4}$ &$-0.87 \times 10^{-2}$
                                 &0.428 &0.084 &0.024 &0.084 \\
           &  &  &  &  &  &    \\
\hline
\end{tabular}
\caption{The same as in Table~1 for $\vartheta_e=15^\circ$. }
\end{table}
\par
Finally, we address the issue problem of the finite acceptance of the 
hadron spectrometers, which necessarily leads to consider the influence of 
the fifth structure function in the measured asymmetry. 
Since the typical values of the vertical angular acceptance are 
$\Delta\phi = \pm 60$ mrad, we have to consider the PC helicity asymmetry 
$\APC$ in the out-of-plane kinematics, just a few degrees 
above and below the electron scattering plane. 
\par
Similarly to the case of the in-plane kinematics, we report in the same 
figure the results of $\APC$ corresponding to a full reaction plane , 
characterizing with positive values of $\tetap$ 
the half-plane $\phi$ and with negative values of $\tetap$ 
the half-plane $180^\circ + \phi$. The results of $\APC$ reported 
in Fig.6 are for the case of backward emitted electrons and for 
$\phi = 2^\circ$ and $4^\circ$. 
In the calculations of $\APC$ we have also included 
meson exchange currents 
of pionic range and the main relativistic corrections (Darwin-Foldy and 
spin-orbit terms as well as the wave function relativistic 
modifications) which have been shown ~\cite{MR90} to be sizeable 
in both the longitudinal-transverse interference structure functions 
$f^{(em)}_{TL}$ and $f^{(em)h}_{TL}$. 
\begin{figure}
\centerline{\psfig{figure=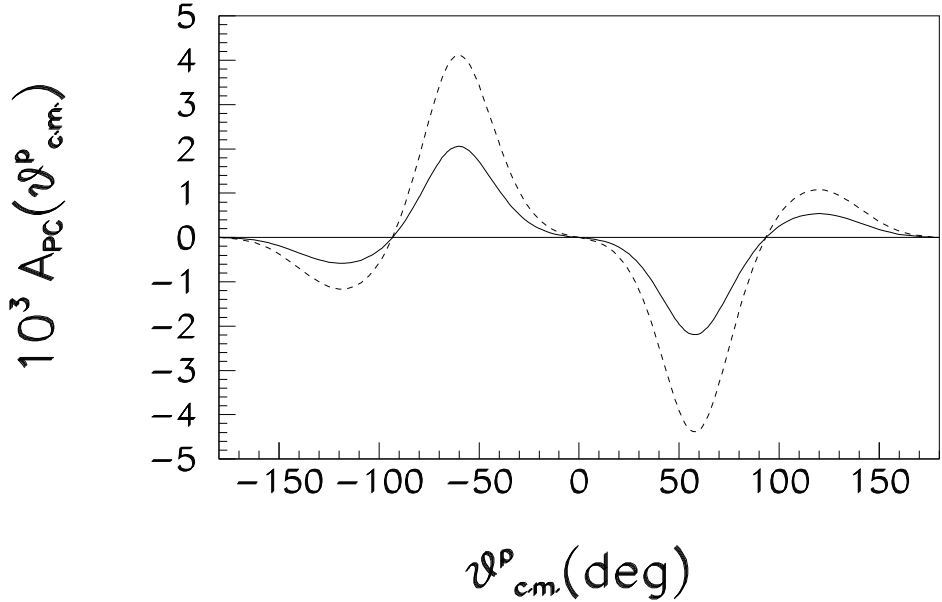,width=14.0cm}} 
\caption{
   Angular distribution of the PC proton asymmetry 
   ${\cal A}_{PC}(\tetacm^p)$ at 
   $Q^2$=0.1$(GeV/c)^2$, $\vartheta_e=160^\circ$, 
   for $\phi=2^\circ$ (full line) and $\phi=4^\circ$ (dashed line). 
   Calculations are with the Paris potential. 
}
\end{figure}
The slight asymmetry of $\APC$ in the two half-planes fully comes 
from the small term proportional to $\cos\phi$ in ${\cal F}$. 
Since $\APC$ is antisymmetric around the electron scattering plane 
the results of $\APC$ in the half-planes $360^\circ-\phi$ and 
$180^\circ-\phi$ follow from those in Fig.6 by a simple change of sign. 
We can see that, apart from the very forward and backward angles, 
the size of $\APC$ is some units of $10^{-3}$, i.e. two order of 
magnitude higher than the PV asymmetry, thus requiring an extremely 
high level of symmetry in the spectrometers in order to make 
negligible the PC contributions to the measured asymmetry. 
\par
Actually, all our considerations of the coincidence PV asymmetry are 
for the strict QE peak, i.e. for the region around $\tetap=0^\circ$ 
where the situation is much more favorable because the fifth structure 
function vanishes at  $\tetap=0^\circ$, as can be seen in Fig.7. 
Here, for typical values ($3^\circ - 4^\circ$) of the horizontal 
angular acceptance of 
the spectrometers, the PC asymmetry drops to some units 
of $10^{-5}$. Therefore, if the spectrometers are symmetrical to 5 
parts in $10^4$, the PC asymmetry should be cancelled at  
the $10^{-8}$ level, thus allowing one to determine the PV asymmetry to 
a few percent. 
\begin{figure}
\centerline{\psfig{figure=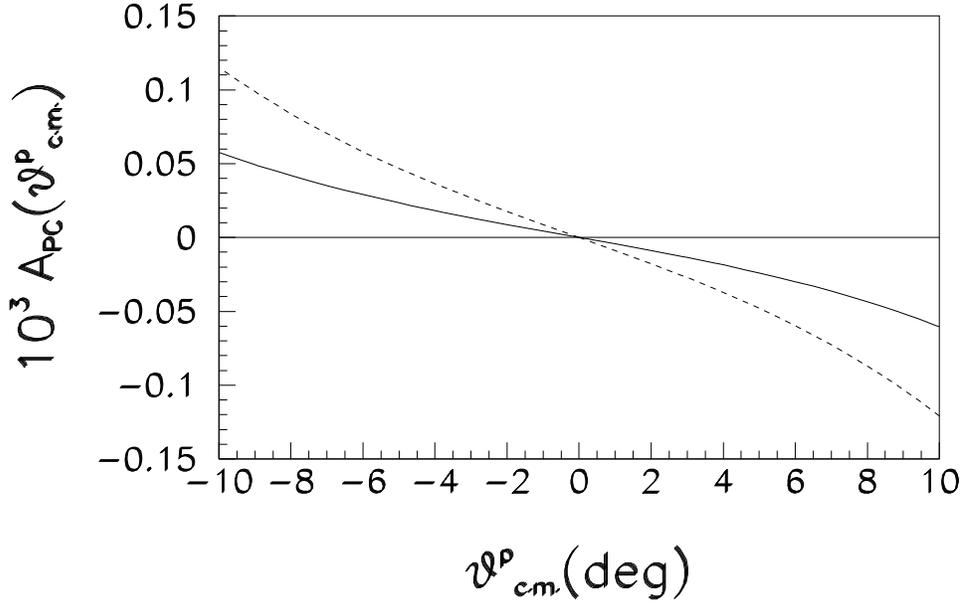,width=14.0cm}} 
\caption{
   The same as in Fig.6 but for the restricted range of the proton 
   emission angle $0^\circ \leq \tetap \leq 10^\circ$. 
}
\end{figure}
\par
The situation is quite similar in the other case considered, i.e. for 
forward emitted electrons. Here the PC asymmetry is one order of 
magnitude smaller than in the previous kinematical case but the PV 
asymmetry is some units of $10^{-6}$ around $\tetap=0^\circ$. 
\section{Conclusions}
\h The aim of this paper was to extend the possible PV observables 
which could be used for an experimental determination of the weak 
form factors of the nucleon. To this end we have considered the 
helicity asymmetry of the 
${\vec e}d$ exclusive cross section in the coplanar geometry which, 
vanishing in a PC theory, directly probes the weak neutral currents. 
\par
First of all, we have derived the general expression of the 
exclusive cross section in the electroweak theory (a result not yet reported 
in the literature to our knowledge). 
From this we have deduced the in-plane helicity asymmetry which depends 
on six structure functions, four of which deriving from the interference 
of the em current and the weak current and two from the interference 
of the em current and the weak axial current. 
We have also given an approximate expression of $\Apteta$ valid 
at $\tetacm=0^\circ$, which allows one to discuss in simple terms the 
importance of the various weak form factors.
\par
Our expectation that the PV exclusive asymmetry should be of interest for 
the determination of the strangeness form factors has been confirmed by actual 
calculations. The point is that the asymmetry of the ${\vec e}d$ exclusive 
cross section in the QE region 
allows one to determine the PV asymmetries of both the 
electron-proton and electron-neutron scattering under the same 
experimental conditions. 
\par
Numerically, we have studied such PV asymmetry in the low $Q^2$ limit in order 
to minimize the impact of the uncertainty on the $Q^2$ dependence of the 
form factors.
\par
We have shown that an experiment with electrons scattered at backward angles 
could allow one to tightly constrain the value of the 
strangeness magnetic moment. 
We have also shown that the asymmetry in the case of forward detected 
electrons is very sensitive to the strangeness radius. However, the precision 
in the extraction of $\rhos$ is rather small because the uncertainties in 
other quantities, and in particular on $\mus$, lead to large errors. 
\par
Finally, we have considered the problem connected with  the finite 
angular acceptance of the spectrometers and with their possible 
asymmetry in the vertical angles, which could lead to include in 
the measured helicity asymmetry some contributions from the PC 
helicity asymmetry. We have shown that 
at the QE peak ($\tetacm=0^\circ$) the PV 
asymmetry can be determined to a few percent if the spectrometers 
are symmetrical to several parts in $10^4$.  
\section{Acknowledgements}
We thank J. Heidenbauer for providing us with the wave functions 
corresponding to the OBEPF potential. 
\par
This work was partly supported by Ministero della Universit\`a e 
della Ricerca Scientifica of Italy.
%
%

%
\end{document}